\documentclass[12pt]{article}
\usepackage{graphicx}

\newsavebox{\sboxpubnumber}
\newsavebox{\sboxpubdate}
\newcommand{\pubdate}[1]{\begin{lrbox}{\sboxpubdate}{#1}\end{lrbox}}
\newcommand{\pubnumber}[1]{\begin{lrbox}{\sboxpubnumber}{\begin{tabular}{l} #1
\\
                 \usebox{\sboxpubdate}
                 \end{tabular}}
                           \end{lrbox}
                           \pubblock}
\newcommand{\Title}[1]{\begin{center} {\Large #1 } \end{center}}
\newcommand{\Author}[1]{\begin{center}{ \sc #1} \end{center}}
\newcommand{\Address}[1]{\begin{center}{ \it #1} \end{center}}

\newcommand{\pubblock}{\rightline{
            \usebox{\sboxpubnumber}}}
\newenvironment{Abstract}{\begin{quotation}  }{\end{quotation}}
\newenvironment{Presented}{\begin{quotation} \begin{center}
             PRESENTED AT\end{center}\bigskip
      \begin{center}\begin{large}}{\end{large}\end{center}
      \end{quotation}}
\newcommand{\Acknowledgements}{\bigskip  \bigskip \begin{center} \begin{large}
             \bf ACKNOWLEDGEMENTS \end{large}\end{center}}

\def\beq{\begin{equation}}
\def\eeq#1{\label{#1}\end{equation}}
\def\eeqn{\end{equation}}

\def\beqa{\begin{eqnarray}}
\def\eeqa#1{\label{#1}\end{eqnarray}}
\def\eeqan{\end{eqnarray}}

\let\bar=\overbar

\def\Dslash{\not{\hbox{\kern-4pt $D$}}}
\def\dslash{\not{\hbox{\kern-2pt $\del$}}}

\def\msb{{\bar{\ssstyle M \kern -1pt S}}}

\begin{document}

\begin{titlepage}
\pubdate{31st January 2002}                    
\pubnumber{~} 

\vfill
\Title{String Cosmology}
\vfill
\Author{Edmund.J.~Copeland}
\Address{Centre for Theoretical Physics, University of Sussex, Falmer, Brighton BN1
9QJ,~~~U.~K.}
\vfill
\begin{Abstract}
We present a brief review of recent advances in string cosmology.
Starting with the Dilaton-Moduli Cosmology (known also as the Pre
Big Bang), we go on to include the effects of axion fields and
address the thorny issue of the Graceful Exit in String Cosmology.
This is followed by a review of density perturbations arising in
string cosmology and we finish with a brief introduction to the
impact moving five branes can have on the Dilaton-Moduli
cosmological solutions.
\end{Abstract}
\vfill
\begin{Presented}
    COSMO-01 \\
    Rovaniemi, Finland, \\
    August 29 -- September 4, 2001
\end{Presented}
\vfill
\end{titlepage}
\def\thefootnote{\fnsymbol{footnote}}
\setcounter{footnote}{0}

\section{Introduction}
String theory, and its most recent incarnation, that of M-theory,
has been accepted by many as the most likely
candidate theory to unify the forces of nature as it
includes General Relativity in a consistent
quantum theory. If it is to play such a pivotal role in particle
physics, it should also include in it all of cosmology. It should
provide the initial conditions for the Universe, perhaps even
explain away the singularity associated with the standard big
bang. It should also provide a mechanism for explaining the
observed density fluctuations, perhaps by providing the inflaton
field or some other mechanism which would lead to inflation.
Should the observations survive the test of time, string theory
should be able to provide a mechanism to explain the current
accelerated expansion of the Universe. In other words, even though
it is strictly a theory which can unify gravity with the
other forces in the very early Universe, for consistency, as a
theory of everything it will have a great deal more to explain. In
this article, we will introduce some of the developments that have
occurred in string cosmology over the past decade or so, initially
basing the discussion on an analyse of the low energy limit of
string theory, and then later extending it to include branes
arising in Heterotic M-theory.

\section{Dilaton-Moduli Cosmology (Pre-Big Bang)}

Strings live in 4+d spacetime dimensions, with the extra $d$
dimensions being compactified. For
homogeneous, four--dimensional cosmologies, where all fields are
uniform on the surfaces of homogeneity, we can consider
the compactification of the
$(4+d)$--dimensional theory on an isotropic $d$--torus. The
radius, or `breathing mode' of the internal space, is then
parameterized by a modulus field, $\beta$, and determines the
volume of the internal dimensions. We can then
assume that the $(4+d)$--dimensional metric is of
the form
\begin{equation}
\label{isotropictorus}
ds^2 =-dt^2 +g_{ij} dx^idx^j +
e^{\sqrt{2/d} \beta} \delta_{ab} dX^a dX^b
\end{equation}
where indices run from $(i,j)=(1, 2, 3)$ and $(a,b)=(4, \ldots ,
3+d)$ and $\delta_{ab}$ is the $d$--dimensional
Kronecker delta. The modulus field $\beta$ is normalized
in such a way that it becomes minimally coupled to gravity
in the Einstein frame.

The low energy action that is commonly used as a starting point
for string cosmology is the four dimensional effective
Neveu-Schwarz- Neveu-Schwarz (NS-NS) action given by:

\begin{equation}
\label{reduced1}
S_*=\int d^4 x \sqrt{|g|} e^{-\varphi} \left[ R +
\left( \nabla \varphi \right)^2 -\frac{1}{2}
\left( \nabla \beta \right)^2 -\frac{1}{2}
e^{2\varphi} \left( \nabla\sigma \right)^2
\right]\, ,
\end{equation}
where $\varphi$ is the effective dilaton in four dimensions, and
$\sigma$ is the pseudo--scalar axion field which is dual to the
fundamental NS--NS three--form
field strength present in string theory, the duality being given by
\begin{equation}
\label{dualHstring}
H^{\mu\nu\lambda} = \epsilon^{\mu\nu\lambda\kappa}
\, e^{\varphi} \, \nabla_{\kappa} \sigma.
\end{equation}

The dimensionally reduced action (\ref{reduced1}) may be viewed as
the prototype action  for string cosmology because it contains
many of the key features common to more general actions.
Cosmological solutions to these actions have been extensively
discussed in the literature -- for a review see \cite{LWC}.
Some of them play a
central role in the pre--big bang inflationary scenario, first
proposed by Veneziano \cite{Veneziano91,GasVen93a}. An important
point can be seen immediately in (\ref{reduced1}) where there is a
non-trivial coupling of the dilaton to the axion field, a coupling
which will play a key role later on when we are investigating the
density perturbations arising in this scenario.

All homogeneous and isotropic external four--dimensional spacetimes can
be described by the Friedmann-Robertson-Walker (FRW) metric.
The general  line element in the string frame can be written as
\begin{equation}
\label{FRW}
ds_4^2 = a^2(\eta) \left\{ -d\eta^2 + d\Omega_\kappa^2 \right\}
 \, ,
\end{equation}
where $a(\eta)$ is the scale factor of the universe, $\eta$
is the conformal time and
$d\Omega_\kappa^2$ is the line element on a 3-space with constant
curvature $\kappa$:
\begin{equation}
\label{constantline}
d\Omega^2_{\kappa}
=d\psi^2 +\left( \frac{\sin \sqrt{\kappa}\psi}{\sqrt{\kappa}}
\right)^2 \left( d\theta^2 +\sin^2  \theta d\varphi^2 \right)
\end{equation}
To be compatible with a homogeneous and isotropic metric,
all fields, including
the pseudo--scalar axion field, must be spatially homogeneous.

The models with vanishing form fields, but time-dependent
dilaton and moduli fields, are known as
{\em dilaton-moduli-vacuum} solutions.
In the Einstein--frame, these solutions may be interpreted
as FRW cosmologies for a stiff perfect fluid,
where the speed of sound equals the speed of light.
The dilaton and moduli fields behave collectively as a massless,
minimally coupled scalar field, and the
scale factor in the Einstein frame is given
by
\begin{equation}
\label{einsteinscalefactor}
\tilde{a} =\tilde{a}_* \sqrt{\frac{\tau}{1+\kappa \tau^2}}
\end{equation}
where $\tilde{a} \equiv e^{-\varphi /2}a$, $\tilde{a}_*$ is
a constant and
we have defined a new time variable:
\begin{equation}
\label{wanmimtime}
\tau \equiv \left\{
\begin{array}{ll}
\kappa^{-1/2} |\tan(\kappa^{1/2}\eta)| & {\rm for}\ \kappa>0 \\
|\eta|  & {\rm for}\ \kappa=0 \\
|\kappa|^{-1/2} |\tanh(|\kappa|^{1/2}\eta)| & {\rm for}\ \kappa<0
\end{array}
\right.
\ .
\end{equation}
The time coordinate $\tau$ diverges at both early and late times in
models which have $\kappa\geq0$, but $\tau\to|\kappa|^{-1/2}$ in
negatively curved models.
There is a curvature singularity at $\eta =0$ with $\tilde{a} =0$
and the model expands away from it
for $\eta >0$ or collapses towards it for $\eta < 0$.
The expanding, closed models recollapse at
$\eta = \pm \pi /2$ and there are no bouncing solutions in
this frame.

\begin{figure}[htb]
    \centering
    \includegraphics[height=9cm]{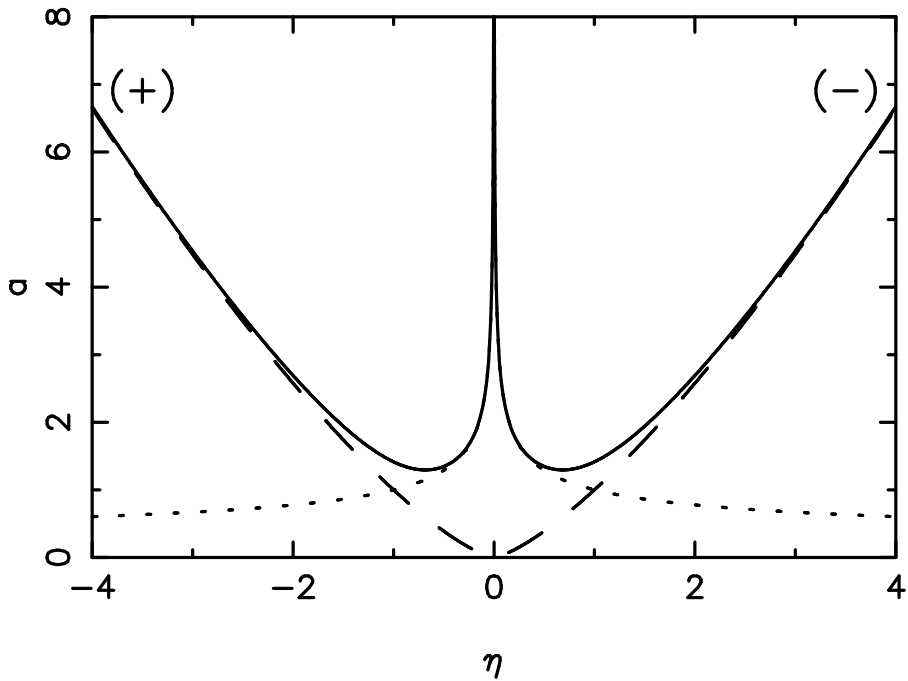}
    \caption{String frame scale factor, $a$, as a function of conformal time, $\eta$, for
flat $\kappa=0$ FRW cosmology in dilaton-vacuum solution in
Eq.~(\ref{dila})
with $\xi_*=0$ (dashed-line), $\xi_*=\pi$ (dotted
line) and dilaton-axion solution in Eq.~(\ref{axia})
with $r=\sqrt{3}$
(solid line). The $(+)$ and $(-)$ branches are defined in
the text.}
    \label{figflatfrw1}
\end{figure}

\begin{figure}[htb]
    \centering
    \includegraphics[height=9cm]{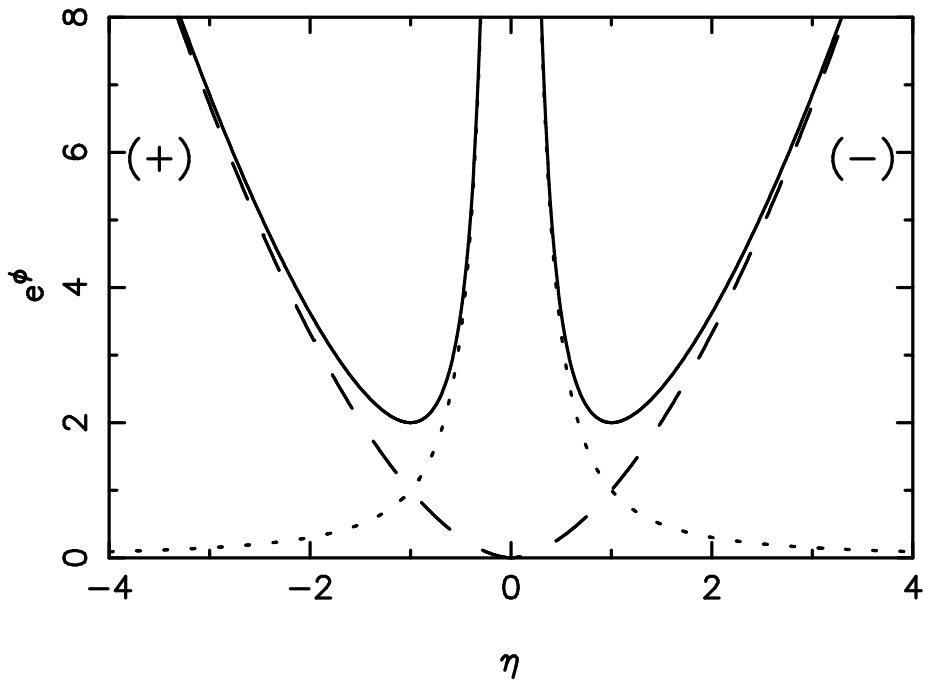}
    \caption{Dilaton, $e^\varphi$, as a function of conformal time, $\eta$, for
flat $\kappa=0$ FRW cosmology in dilaton-vacuum solution in
Eq.~(\ref{dilphi}) with $\xi_*=0$ (dashed-line), $\xi_*=\pi$ (dotted
line) and dilaton-axion solution in Eq.~(\ref{axiphi}) with $r=\sqrt{3}$
(solid line).}
    \label{figflatfrw2}
\end{figure}

The corresponding string frame scale factor,
dilaton and modulus fields are given by the `rolling radii'
solutions \cite{CopLahWan94}

\begin{eqnarray}
\label{dila}
a & = & a_* \sqrt{ {\tau^{1+\sqrt{3}\cos\xi_*} \over 1+\kappa\tau^2} } \, ,
\\
\label{dilphi}
e^\varphi & = & e^{\varphi_*} \tau^{\sqrt{3}\cos\xi_*}
 \, ,\\
\label{dilbeta}
e^\beta & = & e^{\beta_*} \tau^{\sqrt{3}\sin\xi_*} \,
\end{eqnarray}
The integration constant
$\xi_*$ determines the rate of change of the effective dilaton
relative to the volume of the internal dimensions.  Figures
\ref{figflatfrw1} and \ref{figflatfrw2} show the dilaton-vacuum solutions
in flat FRW models when stable compactification has
occurred, so that the volume of the internal space is fixed, with
$\xi_*\ {\rm mod}\ \pi=0$.

The solutions just presented have a
scale factor duality which when applied simultaneously
with time reversal implies that
the Hubble expansion parameter $H \equiv d(\ln a) /dt$
remains invariant,
$H(-t) \to H(t)$, whilst its first derivative
changes sign, $\dot{H} (-t) \to - \dot{H} (t)$.
A decelerating, post--big bang solution -- characterized
by $\dot{a}>0$, $\ddot{a}<0$ and $\dot{H}<0$ --  is mapped onto
a pre--big bang phase of inflationary expansion, since $\ddot{a} /a
= \dot{H}+H^2 >0$. The Hubble radius $H^{-1}$
decreases with increasing time and the expansion is therefore
super-inflationary.
Thus, the pre-big bang cosmology ($\kappa=0$ case in
Eqns.~(\ref{dila}--\ref{dilbeta})) is one that
has a period of super-inflation driven simply by the kinetic energy of the
dilaton and moduli fields \cite{Veneziano91,GasVen93a}.
This is related by duality
to the usual FRW post--big bang
phase. The two branches are separated by a curvature singularity,
however, and it is not clear how the
transition between the pre-- and post--big bang phases
might proceed. This will be the focus of attention in section five.

The solution for a flat ($\kappa=0$) FRW universe
corresponds to the well--known
monotonic power-law, or `rolling radii', solutions.
For $\cos\xi_*<-1/\sqrt{3}$ there is accelerated expansion, i.e.,
inflation, in the string frame for $\eta<0$ and $e^{\varphi} \to 0$ as
$t \to -\infty$, corresponding to the weak coupling regime.
The expansion is an example of `pole--law' inflation \cite{PolSah89,LevFre93}.

The solutions have semi-infinite proper lifetimes. Those starting
from a singularity at $t=0$ for $t\ge 0$ are denoted as the (--)
branch in Ref.~\cite{BruVen94}, while those which approach a
singularity at $t=0$ for $t\leq0$ are referred to as the $(+)$
branch (see figures \ref{figflatfrw1}--\ref{figflatfrw2}). These
$(+/-)$ branches do {\em not} refer to the choice of sign for
$\cos\xi_*$. On either the $(+)$ or $(-)$ branches of the
dilaton-moduli-vacuum cosmologies we have a one-parameter family
of solutions corresponding to the choice of $\xi_*$, which
determines whether $e^\varphi$ goes to zero or infinity as
$t\to0$. These solutions become singular as the conformally
invariant time parameter $\eta \equiv \int dt /a(t) \to 0$ and
there is no way of naively connecting the two branches based
simply on these solutions \cite{BruVen94}.

In the Einstein frame, where the dilaton field is minimally
coupled to gravity, the scale factor given in
Eq.~(\ref{einsteinscalefactor}), becomes
\begin{equation}
\label{Esf}
\tilde{a} = \tilde{a}_*
\left|\eta \right|^{1/2}
\end{equation}
As
$\eta \to 0$ on the (+) branch, the universe is collapsing with
$\tilde{a} \to 0$, and the comoving Hubble length
$|d(\ln\tilde{a})/d\eta|^{-1}=2|\eta|$ is decreasing with time. Thus,
in both frames there is inflation taking place in the sense that a
given comoving scale, which starts arbitrarily far within the
Hubble radius in either conformal frame as $\eta \to - \infty$,
inevitably becomes larger than the Hubble radius in that frame as
$\eta \to 0$. The significance of this is that it means that
perturbations can be produced in the dilaton, graviton and other
matter fields on scales much larger than the present Hubble radius
from quantum fluctuations in flat spacetime at earlier times --
this is a vital property of any inflationary scenario.

For completeness, it is worth mentioning that these
solutions can be extended to include a time-dependent axion field,
$\sigma(t)$, by exploiting the ${\rm SL}(2,R)$ S-duality invariance of
the four--dimensional, NS-NS action \cite{CopLahWan94}.
We now turn our attention to
this fascinating case.

\section{Dilaton-Moduli-Axion Cosmologies}

The cosmologies containing a non--trivial axion field can be
generated immediately due to the global ${\rm SL}(2,R)$ symmetry
of the action (\ref{reduced1}). The resultant solutions
are~\cite{CopLahWan94}:
\begin{eqnarray}
\label{axiphi}
e^\varphi & = & {e^{\varphi_*} \over 2} \left\{
\left({\tau\over\tau_*}\right)^{-r} +
\left({\tau\over\tau_*}\right)^{r}
\right\} \, ,\\
\label{axia}
a^2 & = & \frac{a_*^2}{2(1+\kappa \tau^2)}
\left\{ \left({\tau\over\tau_*}\right)^{1-r}
 + \left({\tau\over\tau_*}\right)^{1+r} \right\} \, ,\\
\label{axibeta}
e^\beta & = & e^{\beta_*} \tau^{s} \, ,\\
\label{axisigma}
\sigma & = & \sigma_*
 \pm e^{-\varphi_*} \left\{ (\tau/\tau_*)^{-r} -
 (\tau/\tau_*)^{r}
\over (\tau/\tau_*)^{-r} +
 (\tau/\tau_*)^{r} \right\} \, ,
\end{eqnarray}
where the exponents are related via
\begin{equation}
r^2+s^2=3 \ ,
\end{equation}
and without loss of generality we may take $r\geq0$.

\begin{figure}[htb]
    \centering
    \includegraphics[height=9cm]{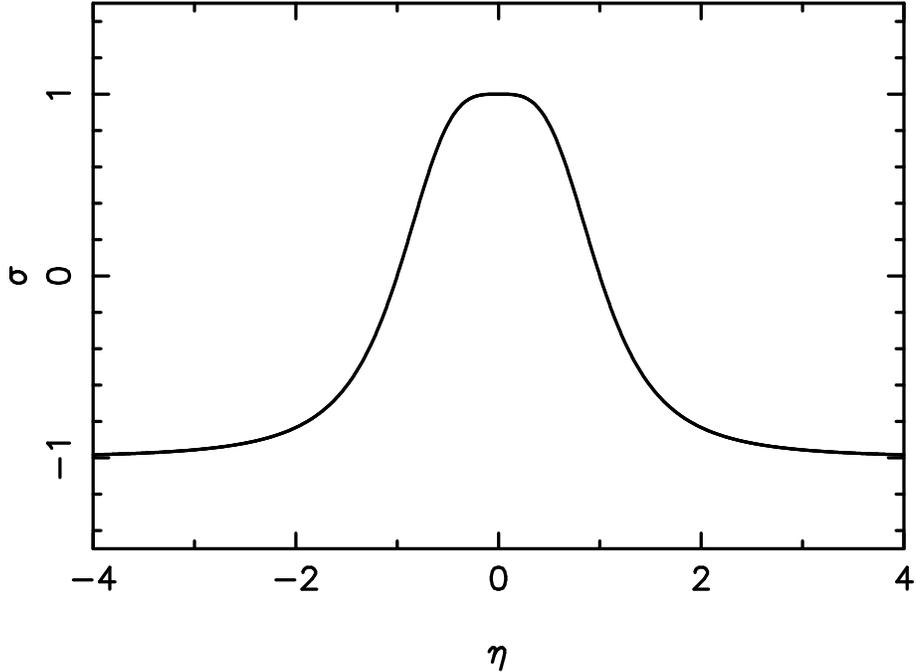}
    \caption{Axion, $\sigma$, as a function of conformal time, $\eta$, for flat
$\kappa=0$ FRW cosmology in dilaton-axion solution in Eq.~(\ref{axisigma})
with $r=\sqrt{3}$ (solid line).}
    \label{figNSNSsigma}
\end{figure}

In all cases, the dynamics of the axion field places a {\em lower}
bound on the value of the dilaton field, $\varphi\geq\varphi_*$. In
so doing, the axion smoothly interpolates between two
dilaton--moduli--vacuum solutions, where its dynamical influence
asymptotically becomes negligible.  The effects of
time--dependent axion
solutions for the scale-factor and dilaton
are plotted in Figure2 \ref{figflatfrw1}
and \ref{figflatfrw2}
for the flat FRW model when the modulus field is trivial
$(s=0)$. When the internal space is static, it is seen that the string
frame scale factors exhibit a bounce.
However we still have a curvature singularity in the
Einstein frame as $\tau\to0$.
The actual time-dependent axion solutions
is shown in Figure \ref{figNSNSsigma}.

The spatially flat solutions reduce to the power law,
dilaton--moduli--vacuum solution given in
Eqs.~(\ref{dila}--\ref{dilbeta}) at early and late times.  When
$\eta\to\pm\infty$ the solution approaches the vacuum solution
with $\sqrt{3}\cos \xi_* =+r$, while as $\eta\to0$ the solution
approaches the $\sqrt{3}\cos \xi_* =-r$ solution. Thus, the axion
solution interpolates between two vacuum solutions related by an
S-duality transformation $\varphi\to-\varphi$.  When the internal
space is static the scale factor in the string frame is of the
form $a \propto t^{1/\sqrt{3}}$ as $\eta\to\pm\infty$, while as
$\eta\to0$ the solution becomes $a \propto t^{-1/\sqrt{3}}$. These
two vacuum solutions are thus related by a scale factor duality
that inverts the spatial volume of the universe. This asymptotic
approach to dilaton--moduli--vacuum solutions at early and late
times will lead to a particularly simple form for the
semi-classical perturbation spectra that is independent of the
intermediate evolution. However, there is a down side to these
solutions from the standpoint of pre big bang cosmologies. As
$\eta\to\pm\infty$ and as $\eta\to 0$ the solution approaches the
strong coupling regime where $e^\varphi \to \infty$. Thus there is
no weak coupling limit, the axion interpolates between two
strong coupling vacuum solutions. We will shortly see how a
similar affect arises when we include a moving brane in the
dilaton-moduli picture, as it too mimics the behaviour of
a non-minimally coupled
axion field.

The overall dynamical effect of the axion field is negligible except near
$\tau\approx \tau_*$, when it leads to a bounce in the dilaton field.
Within the context of M--theory cosmology, the radius of the eleventh
dimension is related to the dilaton by
$r_{11} \propto e^{\varphi /3}$ when the modulus
field is fixed. This bound on the dilaton may therefore be reinterpreted
as a lower bound on the size of the eleventh dimension.

\section{Fine tuning issues}

The question over the viability of the initial conditions required in
the pre Big Bang scenario has been a cause for many an argument both in print
and in person. Since both $\dot{H}$ and $\dot{\varphi}$ are positive in the
pre--big bang phase, the initial values for these
parameters must be {\em very small}. This raises a number of important
issues concerning fine--tuning in the pre--big bang
scenario~\cite{TurWei97,ClaLidTav98,KalLinBou99,MahOnoVen98,Veneziano97,
BuoMeiUng98,BuoDamVen99}. There needs to be enough
inflation in a homogeneous patch in order
to solve the horizon and flatness problems which means that the
dilaton driven inflation must survive for a
sufficiently long period of
time. This is not as trivial as it may appear, however,  since the period of
inflation is limited by a number of factors.

The fundamental postulate of the scenario  is that
the initial data for inflation lies well within the
perturbative regime of string theory,
where the curvature and coupling are very
small \cite{GasVen93a}. Inflation then proceeds
for sufficiently homogeneous initial conditions
\cite{Veneziano97,BuoMeiUng98},
where time derivatives are dominant with respect to
spatial gradients, and the universe evolves into a high curvature and
strongly--coupled regime.
Thus, the pre--big bang initial state should correspond to a
cold, empty and flat vacuum state.
Initial the universe would have been huge relative
to the quantum scale and hence should have been well described by
classical solutions to the string effective action. This should be
compared to the initial state which describes the standard hot big
bang, namely a dense, hot, and highly curved region of spacetime.
This is quite a contrast and
a primary goal of pre--big bang  cosmology must be to develop a
mechanism for smoothly
connecting these two regions, since we believe that the standard big
bang model provides a very good representation of the current evolution of
the universe.

Our present observable universe
appears very nearly homogeneous on sufficiently large scales.
In the standard, hot big bang model,
it corresponded to a region at the Planck time that was $10^{30}$ times larger
than the horizon size, $l_{\rm Pl}$.  This
may be viewed as an initial condition in the big bang model or as a final
condition for inflation.
It implies that the  comoving Hubble radius, $1/(aH)$,
must decrease during
inflation by a factor of at least $10^{30}$ if the horizon problem is to
be solved. For
a power law expansion, this implies that
\begin{equation}
\label{bound}
\left| \frac{\eta_f}{\eta_i} \right| \le 10^{-30}
\end{equation}
where subscripts $i$ and $f$
denote values at the onset and end of inflation, respectively.

In the pre--big bang scenario, Eq. (\ref{dilphi})
implies that the dilaton grows as $e^{\varphi} \propto |
\eta |^{-\sqrt{3}}$, and since at the start of the post--big bang epoch, the
string coupling, $g_s =e^{\varphi /2}$, should be of order unity,
the bound (\ref{bound}) implies that the initial value of the
string coupling is strongly constrained, $g_{s,i} \le 10^{-26}$.
Turner and Weinberg interpret this constraint as a severe fine--tuning
problem in the scenario, because inflation in the string frame can
be delayed by the effects of spatial curvature
\cite{TurWei97}. It was shown by Clancy, Lidsey and Tavakol that the
bounds are further tightened when spatial anisotropy is introduced, actually
preventing  pre--big bang inflation
from occurring \cite{ClaLidTav98}. Moreover, as we have seen the dynamics of
the NS--NS axion field also places a lower bound on the
allowed range of values that the string coupling may
take \cite{CopLahWan94}.
In the standard inflationary scenario,
where the expansion is quasi--exponential, the Hubble radius is approximately
constant and $a \propto (-\eta )^{-1}$. Thus, the homogeneous
region grows by a
factor of $|\eta_i /\eta_f |$ as inflation proceeds. During a pre--big bang
epoch, however, $a \propto ( -\eta )^{-1/1+\sqrt{3}}$ and
the increase in the size of a homogeneous region is
reduced by a factor of at least $10^{30 \sqrt{3}/(1+\sqrt{3})} \approx
10^{19}$ relative to that of the standard inflation
scenario. This implies that the initial
size of the homogeneous region should exceed $10^{19}$ in string units if
pre--big bang inflation is to be successful in solving the problems
of the big bang model
\cite{Veneziano91,KalLinBou99}. The occurrence of such a large
number was cited by
Kaloper, Linde and Bousso as a serious problem of the pre--big
bang scenario, because it implies that the universe must already
have been large and smooth by the time inflation began \cite{KalLinBou99}.

On the other hand, Gasperini has emphasized
that the initial homogeneous region of the pre--big bang universe is
not larger than the horizon even though it is large relative
to the string/Planck scale \cite{Gasperini99}.
The question that then arises when discussing the naturalness, or
otherwise, of the above initial conditions is what
is the basic unit of length that should be employed.
At present, this question has not been addressed in detail.

Veneziano and collaborators conjectured that pre--big bang
inflation generically evolves out of an initial state that
approaches the Milne universe in the semi--infinite past,
$t \rightarrow -\infty$ \cite{Veneziano97,BuoMeiUng98}.
The Milne universe may be mapped onto the future (or past) light cone of the
origin of Minkowski spacetime and therefore corresponds to a
non--standard representation of the string
perturbative vacuum. The proposal was that the Milne
background represents an early time attractor, with a large
measure in the space of initial data. If so, this
would provide strong justification for the postulate that
inflation begins in the weak coupling and curvature regimes
and would render the pre-big bang assumptions regarding the
initial states as `natural'. However,
Clancy {\em et al.} took a critical look at this conjecture and
argued that the Milne universe is an unlikely
past attractor for the pre--big bang scenario \cite{ClaLidTav99}. They
suggested that plane wave backgrounds represent a more
generic initial state for the universe \cite{ClaLidTav98}.
Buonanno, Damour and Veneziano have subsequently proposed that
the initial state of the pre--big bang universe should correspond to
an ensemble of gravitational and dilatonic waves \cite{BuoDamVen99}.
They refer to this as the state of `asymptotic past triviality'.
When viewed in the Einstein
frame these waves undergo collapse when certain conditions are
satisfied. In the string frame, these gravitationally unstable
areas expand into homogeneous regions on large scales.

To conclude this Section, it is clear that the question
of initial conditions in the pre--big bang scenario
is currently unresolved. We turn our attention now to another
unresolved problem for the scenario -- the Graceful Exit.

\section{The Graceful Exit}

We have seen how in the pre Big Bang scenario, the Universe
expands from a weak coupling, low curvature regime in the infinite
past, enters a period of inflation driven by the kinetic energy
associated with the massless fields present, before approaching
the strong coupling regime as the string scale is reached. There
is then a branch change to a new class of solutions, corresponding
to a post big bang decelerating Friedman-Robertson-Walker era. In
such a scenario, the Universe appears to emerge because of the
gravitational instability of the generic string vacua --  a very
appealing picture, the weak coupling, low curvature regime is a
natural starting point to use the low energy string effective
action. However, how is the branch change achieved without hitting
the inevitable looking curvature singularity associated with the
strong coupling regime? The simplest version of the evolution of
the Universe in the pre-big bang scenario inevitably leads to a
period characterised by an unbounded curvature. The current
philosophy is to include higher-order corrections to the string
effective action. These include both classical finite size effects
of the strings ($\alpha'$ corrections arising in higher order
derivatives), and quantum string loop corrections ($g_s$
corrections). The list of authors who have worked in this area is
too great to mention here, for a detailed list see
\cite{LWC,Gweb}. A series of key papers were written by
Brustein and Madden, in which they
demonstrated that it is possible to include such terms and
successfully have an exit from one branch to the other
\cite{BruMad97,BruMad98}. More recently this approach has been
generalised by including combinations of classical
and quantum corrections \cite{CarCopMad00}.
Brustein and Madden \cite{BruMad97,BruMad98} made use
of the result that classical corrections can stabilize a
high curvature string phase while the evolution is still in the weakly
coupled regime\cite{GasMagVen97}.
The crucial new ingredient that they added was
the inclusion of terms of the type that may result from quantum
corrections to the string effective action and
which induce violation
of the null energy condition (NEC -- The Null Energy Condition
is satisfied if $\rho + p \ge 0$, where $\rho$ and $p$
represent the effective energy density and pressure of the additional
sources). Such extra terms mean that evolution towards a
decelerated FRW phase is possible. Of course this violation of the
null energy condition can not continue indefinitely, and eventually it
needs to be turned off in order to stabilise the dilaton at a fixed
value, perhaps by capture in a potential minimum or by radiation
production -- another problem for string theory!

The analysis of \cite{BruMad97} resulted in a set of
necessary conditions on the evolution in terms of the Hubble
parameters $H_S$ in the string frame, $H_E$ in the Einstein frame
and the dilaton $\varphi$, where they are related by $H_E = e^{\varphi/2}
(H_S-\frac{1}{2} \dot \varphi)$. The conditions were:

\begin{itemize}

\item Initial
conditions of a (+) branch and $H_S,\dot\varphi>0$ require $H_E<0$.

\item
A branch change from (+) to $(-)$ has to occur while $H_E<0$.

\item A
successful escape and exit completion requires NEC violation
accompanied by a bounce in the Einstein frame after the branch change
has occurred, ending up with $H_E>0$.

\item Further evolution is
required to bring about a radiation dominated era in which the dilaton
effectively decouples from the ``matter" sources.

\end{itemize}
In the
work of \cite{BruMad98}, employing both types of string
inspired corrections, the authors made use of the known
fact \cite{GasMagVen97} that $\alpha'$ corrections
created an attractive fixed point for a wide range of initial
conditions which stabilized the evolution in a high
curvature regime with linearly growing dilaton.  This then caused the
evolution to undergo a branch change, all of this occurring for small
values of the dilaton (weak coupling), so the quantum corrections could
be ignored.  However, the linearly growing dilaton meant that the
quantum corrections eventually become important. Brustein and Madden
employed these to induce NEC violation and allow the universe to
escape the fixed point and complete the transition to a decelerated
FRW evolution. As an explicit example in \cite{CarCopMad00} we consider
a string theory motivated example where we include a number of
higher derivative $\alpha'$ terms.
Our starting point is the minimal $4-$dimensional string effective
action:
\begin{eqnarray}
\Gamma^{(0)} &=& \frac{1}{\alpha'}\int d^{4}x \sqrt{-g} {\cal L}^{(0)} \nonumber
\\
 &=& \frac{1}{\alpha'}\int d^{4}x \sqrt{-g} e^{-\varphi} \Bigl\{ R +
 (\partial_{\mu} \varphi)^2 \Bigr\}.
\label{Low}
\end{eqnarray}
By low-energy tree-level effective action, we mean
that the string is propagating in a background of small curvature and the
fields are weakly coupled. However, the evolution from the pre-big bang era to
the present is understood to be characterised by a regime of growing couplings
and curvature. This means that the Universe will have to evolve through a phase
when the field equations of this effective action are no longer valid. Hence,
the low-energy dynamical description has to be supplemented by corrections in
order to reliably describe the transition regime.

The finite size of the string will have an impact on the
evolution of the scale factor when the curvature of the Universe reaches a
critical level, corresponding to the string length scale
$\lambda_{S} \sim \sqrt{\alpha'}$ (fixed in the string frame), and such
corrections
are
expected to stabilise the growth of the curvature into a de-Sitter like  regime
of constant curvature and linearly growing dilaton \cite{GasMagVen97}.
Eventually the dilaton will play a major role, and since the loop expansion is
governed by powers of the string coupling parameter $g_{S} = e^{\varphi}$,
these quantum corrections will modify dramatically the evolution when we reach
the strong coupling region \cite{BruMad97,BruMad98}. This should
correspond to the stage when the Universe completes a smooth transition to the
post-big bang branch, characterised by a fixed value of the dilaton and a
decelerating FRW expansion. One of the unresolved issues of the transition
concerns whether or not the actual exit takes place at large coupling,
$e^{\varphi} \ge 1$. If it occurred whilst the coupling was still
small,
then we would be happy to use the perturbative corrections we are adopting.

The type of corrections we consider involve truncations of the
classical action at order $\alpha'$. The most general
form for a correction to the string action up to fourth-order in
derivatives has been presented in refs
\cite{Meissner,Kaloper2}:
\begin{eqnarray}
\Gamma^{(C)} &=& \frac{1}{\alpha'}\int d^{4}x \sqrt{-g} {\cal L}^{(C)} \nonumber
\\
&=&  -\frac{1}{4}\int d^{4}x \sqrt{-g} e^{-\varphi} \Bigl\{ a R^{2}_{GB} + b
\nabla^2 \varphi (\partial_{\mu} \varphi)^{2} \nonumber \\
&&\hspace{0.6cm} + c \bigl\{ R^{\mu\nu}-\frac{1}{2}g^{\mu\nu}R\bigr\}
\partial_{\mu} \varphi \partial_{\nu} \varphi + d (\partial_{\mu} \varphi)^{4} \Bigr\}.
\label{Class}
\end{eqnarray}
$R^{2}_{GB} = R_{\mu\nu\lambda\rho}
R^{\mu\nu\lambda\rho} -4 R_{\mu\nu}R^{\mu\nu} + R^2$ is the Gauss-Bonnet
combination which guarantees the absence of higher derivatives. In fixing the
different parameters in this action we require that it reproduces the
usual string scattering amplitudes \cite{Metsaev}.
This constrains the coefficient of $R_{\mu\nu\lambda\rho}^2$ with the
result that the pre-factor for the Gauss-Bonnet term has to be $a=-1$.
But the Lagrangian can still be shifted by field redefinitions which
preserve the on-shell amplitudes, leaving the three remaining coefficients
of the classical correction
satisfying the constraint
\begin{equation} \label{constraint}
2 ( b+c) + d = - 16 a.
\end{equation}

There is as yet no definitive calculation of the full loop
expansion of string theory. This is of course a big problem if we
want to try and include quantum effects in analysing the graceful
exit issue. The best we can do, is to propose plausible terms that
we hope are representative of the actual terms that will
eventually make up the loop corrections. We believe that the
string coupling $g_{S}$ actually controls the importance of
string-loop corrections, so as a first approximation to the loop
corrections we multiply each term of the classical correction by a
suitable power of the string coupling \cite{BruMad97,BruMad98}.
When loop corrections are included, we then have an effective
Lagrangian given by
\begin{equation}
{\cal L} = {\cal L}^{(0)} + {\cal L}^{(C)} + A e^{\varphi} {\cal L}^{(C)} +
B e^{2\varphi} {\cal L}^{(C)},
\label{efflag}
\end{equation}
where ${\cal L}^{(0)}$ is given in Eq.~(\ref{Low}) and
${\cal L}^{(C)}$ given in Eq.~(\ref{Class}). The constant
parameters $A$ and $B$ actually control the onset of the loop corrections.

\begin{figure}[htb]
\centering
\includegraphics[height=9cm]{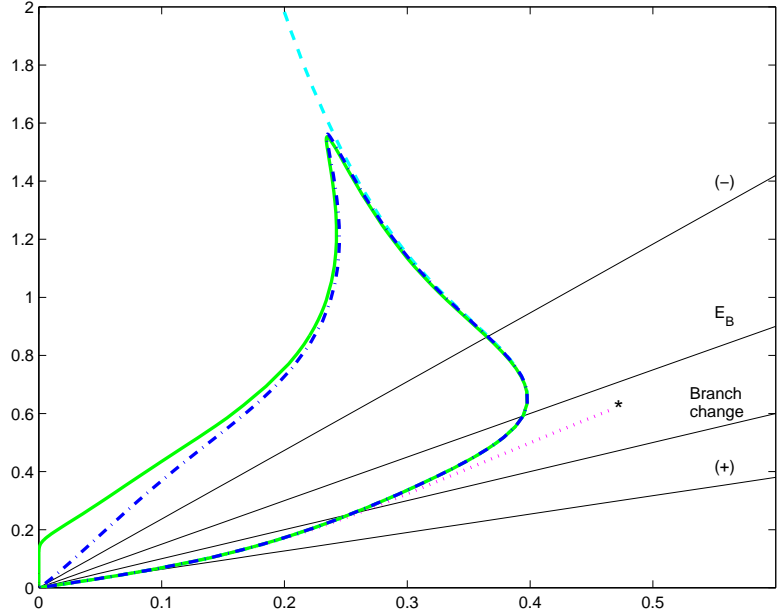}
\caption{Hubble expansion in the S-frame as a function of the
dilaton for a successful exit with
$a=-1$, $b=c=0$ and $d=16$. The y-axis
corresponds to $H$, and the x-axis to $2\dot{\varphi}/3$. The
initial conditions for the simulations have been set with respect
to the lowest-order analytical solutions at $t_{S} = -1000$. The
straight black lines describe the bounds quoted in Section II. The
dotted magenta line shows the impact of the classical correction
due to the finite size of the string. A $*$ denotes the fixed
point. The contribution of the one-loop expansion is traced with a
dashed cyan line ($A=4$). The dash-dotted blue line represents the
incorporation of the two-loop correction without the Gauss-Bonnet
combination ($B=-0.1$). Finally, the green plain line introduces
radiation with $\Gamma_{\varphi}=0.08$ and stabilises the
dilaton.} \label{fig1}
\end{figure}

Not surprisingly the field equations need to be solved numerically, but this can be
done and the solutions are very encouraging as they show
there exists a large class of parameters for
which successful graceful exits are obtained \cite{CarCopMad00}. For example
the natural setting $b=c=0$ leads to the well-known form which has given rise to
most of the studies on corrections to the low-energy picture. In references
\cite{GasMagVen97,BruMad97}, the authors demonstrated that this minimal
classical correction regularises the singular behaviour of the low-energy
pre-big bang scenario. It drives the evolution to a fixed point of bounded
curvature with a linearly growing dilaton (the star in Figure~\ref{fig1} --
which agrees with the results of \cite{GasMagVen97,BruMad97}),
suggesting that quantum loop
corrections -known to allow a violation of the null energy condition
$(p+\rho<0)$- would permit the crossing of the Einstein bounce
to the FRW decelerated expansion in
the post-big bang era. Indeed, the addition of loop corrections
leads to a $(-)$ FRW-branch as pictured in Figure~\ref{fig1}. However, we still
have to freeze the growth of the dilaton. Following
\cite{BruMad97}, we introduce by hand a particle creation term of the form
$\Gamma_{\varphi} \dot{\varphi}$, where $\Gamma_{\varphi}$ is the decay width of the
$\varphi$ particle, in the equation of motion of the dilaton field and
then coupling it to a fluid with the equation of state of radiation in
such a way as to conserve energy overall.
This allows us to stabilise the dilaton in the post-big bang
era with a decreasing Hubble rate,
similar to the usual radiation dominated FRW cosmology. We should point out though, that although it
is possible to have a successful exit, it is not so easy to ensure that the exit takes place in
a weakly coupled regime, and typically we found that as the exit was approached
$\varphi_{\rm final} \sim 0.1 -- 0.3$. Thus it is fair to say that although great progress
has been made on the question of Graceful Exit in string cosmology, it remains a problem
in search of the full solution. It is a fascinating problem, and not surprisingly
alternative prescriptions which aim to address
this issue have recently been proposed, involving colliding branes
\cite{ekpyrotic} and Cyclic universes \cite{cyclic}.
We now turn our attention to the observational consequences
of string cosmology, in particular the generation of the observed cosmic microwave background
radiation.

\section{Density perturbations in String Cosmology}

We have to consider
inhomogeneous perturbations that may be generated due to vacuum
fluctuations, and follow the formalism pioneered by Mukhanov and collaborators
\cite{Mukhanov88,MukFelBra92}.
During a period of accelerated expansion the comoving Hubble length,
$|d(\ln a)/d\eta|^{-1}$, decreases and vacuum fluctuations
which are assumed to start in the flat-spacetime vacuum state may be
stretched up to exponentially large scales. The precise form of the
spectrum depends on the expansion of the homogeneous background and
the couplings between the fields.
The comoving Hubble length,
$|d(\ln\tilde{a})/d\eta|^{-1}=2|\eta|$, does indeed decrease in the
Einstein frame during the contracting phase when $\eta<0$. Because the
dilaton, moduli fields and graviton are minimally coupled to this
metric, this ensures that small-scale vacuum fluctuations will
eventually be stretched beyond the comoving Hubble scale during this epoch.

As we remarked earlier, the axion field is taken to be a constant in
the classical pre-big bang solutions.  However, even when the
background axion field is set to a constant, there will inevitably be
quantum fluctuations in this field.  We will see that these
fluctuations can not be neglected and, moreover, that
they are vital if the pre-big bang scenario is to have
any chance of generating the observed density perturbations.

In the Einstein frame, the first-order perturbed line element can
be written as
\begin{equation}
\label{dds}
d\widetilde{s}^2  =  \widetilde{a}^2(\eta)
 \left\{ -(1+2\widetilde{A})d\eta^2
 + 2\widetilde{B}_{,i} d\eta dx^i
+ \left[\delta_{ij} + h_{ij}\right] dx^i dx^j \right\} \, ,
\end{equation}
where $\widetilde{A}$ and $\widetilde{B}$
are scalar perturbations and $h_{ij}$ is a tensor
perturbation.

\subsection{Scalar metric perturbations}

First of all we consider the evolution of linear metric perturbations
about the four-dimensional spatially flat dilaton-moduli-vacuum
solutions given in Eqs.~(\ref{dila}--\ref{dilbeta}). Considering a
single Fourier mode, with comoving wavenumber $k$, the perturbed
Einstein equations yield the evolution equation
\begin{equation}
\label{Aeom}
\widetilde{A}'' + 2\widetilde{h}\widetilde{A}' + k^2 \widetilde{A}
  =  0 \, ,
\end{equation}
plus the constraint
\begin{equation}
\label{AofB}
\widetilde{A} = - ( \widetilde{B}' +2\widetilde{h}\widetilde{B} ) \, ,
\end{equation}
where $\widetilde{h}$ is the Hubble parameter in the Einstein frame derived from
Eq.~(\ref{Esf}), and $\widetilde{A}' \equiv \frac{d\widetilde{A}}{d \eta}$.
In the spatially flat gauge we have the simplification that the
evolution equation for the scalar metric perturbation, Eq.~(\ref{Aeom}),
is independent of the evolution of the different
massless scalar fields (dilaton, axion and moduli), although they will
still be related by the constraint
\begin{equation}
\label{pBBAcon}
\widetilde{A} = {\varphi'\over4\widetilde{h}} \, \delta\varphi
  +{\beta'\over4\widetilde{h}} \, \delta\beta \, ,
\end{equation}
where $\delta\varphi$ and $\delta\beta$ are the perturbations
in $\varphi$ and $\beta$ respectively. To first-order, the metric
perturbation, $\widetilde{A}$, is determined solely by the dilaton and
moduli field perturbations, although its
evolution is dependent only upon the
Einstein frame scale factor, $\widetilde{a}(\eta)$, given by
Eq.~(\ref{Esf}), which in turn is determined solely by the stiff fluid
equation of state for the homogeneous fields in the Einstein frame.

One of the most useful quantities we can calculate
is the curvature perturbation on uniform energy density hypersurfaces (as
$k\eta\to0$). It is commonly denoted by $\zeta$~\cite{BarSteTur83}and
in the Einstein frame, we obtain
\begin{equation}
\zeta = {\widetilde{A}\over3} \, ,
\end{equation}
in any dilaton--moduli--vacuum or dilaton--moduli--axion cosmology
\cite{BruGasGio95,CopEasWan97}.

The significance of $\zeta$ is that in an expanding universe it
becomes constant on scales much larger than the Hubble scale
($|k\eta|\ll1$) for purely adiabatic perturbations.
In single-field inflation models
this allows one to compute the density perturbation at late times,
during the matter or radiation dominated eras, by equating $\zeta$ at
``re-entry'' ($k=\widetilde{a}\widetilde{H}$) with that at horizon
crossing during inflation.  To calculate $\zeta$, hence
the density perturbations induced in the pre-big bang
scenario we can either use the vacuum
fluctuations for the canonically normalised field at early times/small
scales (as $k\eta\to-\infty$) or use the amplitude of the scalar field
perturbation spectra to normalise the
solution for $\widetilde{A}$. This yields, (after some work),
the curvature perturbation spectrum on large scales/late times (as
$k\eta\to0$):
\begin{equation}
\label{Pzeta}
\label{Aspectrum}
{\cal P}_{\zeta} = {8\over\pi^2}
 l_{\rm Pl}^2\widetilde{H}^2 (-k\eta)^3[\ln(-k\eta)]^2
 \, ,
\end{equation}
where $l_{\rm Pl}$ is the Planck length in the Einstein frame and remains
fixed throughout.
The scalar metric perturbations become large on superhorizon
scales ($|k\eta|<1$) only near the Planck era, $\widetilde{H}^2\sim
l_{\rm Pl}^{-2}$.

The spectral index of the curvature perturbation spectrum is
conventionally given as~\cite{LidLyt93}
\begin{equation}
\label{zetaspecindex}
n \equiv 1+ {d\ln{\cal P}_{\zeta} \over d\ln k}
\end{equation}
where $n=1$ corresponds to the classic Harrison-Zel'dovich spectrum
for adiabatic density perturbations favoured by most models of
structure formation in our universe. By contrast the pre--big bang era
leads to a spectrum of curvature perturbations with $n=4$.
Such a steeply tilted spectrum of
metric perturbations implies that there would be effectively no primordial
metric perturbations on large (super-galactic) scales in our present
universe if the post-Big bang era began close to the Planck scale.
Fortunately, as we shall see later, the presence of the axion field could
provide an alternative spectrum of perturbations more suitable as a
source of large-scale structure.
The pre-big bang scenario is not so straightforward as in the
single field inflation case, because the full low-energy string
effective action possesses many fields which can lead to non-adiabatic
perturbations. This implies that density perturbations at late times
may not be simply related to $\zeta$ alone, but may also be dependent
upon fluctuations in other fields.

\subsection{Tensor metric perturbations}

The gravitational wave perturbations, $h_{ij}$, are both gauge and
conformally invariant. They decouple from the scalar perturbations in
the Einstein frame to give a simple evolution equation for each
Fourier mode
\begin{equation}
\label{heom}
h_k'' + 2\widetilde{h}\, h_k' + k^2 h_k  = 0 \, .
\end{equation}
This is exactly the same as the equation of motion for the scalar
perturbation given in Eq.~(\ref{Aeom}) and has the same
growing mode in the long wavelength ($|k\eta|\to0$) limit given by
Eq.~(\ref{Pzeta}).
The spectrum depends solely on the dynamics of the scale factor in the
Einstein frame given in Eq.~(\ref{Esf}), which remains the
same regardless of the time-dependence of the different dilaton,
moduli or axion fields. It leads to a spectrum of primordial
gravitational waves steeply growing on short scales, with a spectral
index
$n_T=3$~\cite{GasVen93a},
in contrast to conventional inflation models which require
$n_T<0$~\cite{LidLyt93}. The graviton spectrum appears to be a robust
and distinctive prediction of any pre-big bang type evolution based on
the low-energy string effective action, although recently in the non-singular
model of section 5, we have
demonstrated how passing through the string phase could lead to
a slight shift in the tilt closer to $n_T \sim 2$ \cite{CarCopGas}

\subsection{Dilaton--Moduli--Axion Perturbation Spectra}

We will now consider inhomogeneous linear perturbations in the fields
about a homogeneous background given by \cite{CopEasWan97,CopLidWan97}
\begin{equation}
\varphi = \varphi(\eta) + \delta \varphi ({\bf x},\eta), ~~~\sigma = \sigma(\eta)
+ \delta \sigma ({\bf x},\eta), ~~~\beta = \beta (\eta) + \delta \beta
({\bf x},\eta) \ .
\end{equation}
The perturbations can be re-expressed as a Fourier series in terms of
Fourier modes with comoving wavenumber $k$.
Considering the production of dilaton, moduli
and axion perturbations during a pre-big bang evolution where the
background axion field is constant, $\sigma'=0$,
the evolution of the homogeneous background fields are
given in Eqs.~(\ref{dilphi}--\ref{dilbeta}).
The dilaton and
moduli fields both evolve as minimally coupled massless fields
in the Einstein frame. In particular, the dilaton perturbations are
decoupled from the axion perturbations and the equations of motion in
the spatially flat gauge become
\begin{eqnarray}
\label{pBBdphieom}
\delta\varphi'' + 2\widetilde{h}\delta\varphi' + k^2\delta\varphi
 & = & 0 \, , \\
\label{pBBdbetaeom}
\delta\beta'' +2\widetilde{h}\delta\beta' +k^2\delta\beta
 & = & 0 \, ,\\
\label{pBBdsigmaeom}
\delta\sigma''+2\widetilde{h}\delta\sigma' + k^2\delta\sigma
 & = & - 2\varphi'\delta\sigma' \, ,
\end{eqnarray}
Note that these evolution equations for the scalar field perturbations
defined in the spatially flat gauge are automatically decoupled from
the metric perturbations, although as we have said they are still related to the
scalar metric perturbation, $\widetilde{A}$ through Eq.~(\ref{pBBAcon}).

On the $(+)$ branch, i.e., when $\eta<0$, we can normalise modes at
early times, $\eta\to-\infty$, where all the modes are far inside the
Hubble scale, $k\gg|\eta|^{-1}$, and can be assumed to be in
the flat-spacetime vacuum. Whereas in conventional
inflation where we have to assume that this result for a quantum field in a
classical background holds at the Planck scale, in this case the
normalisation is done in the zero-curvature limit in the infinite
past. Just as in conventional inflation, this produces perturbations
on scales far outside the horizon, $k\ll|\eta|^{-1}$, at late times,
$\eta\to0^-$.

Conversely, the solution for the $(-)$ branch with $\eta>0$ is dependent
upon the initial state of modes far outside the horizon,
$k\ll|\eta|^{-1}$, at early times where $\eta\to0$. The role of a period
of inflation, or of the pre-big bang $(+)$ branch, is precisely to set
up this initial state which otherwise appears as a mysterious initial
condition in the conventional (non-inflationary) big bang model.

The power spectrum for perturbations is commonly denoted by
\begin{equation}
{\cal P}_{\delta x} \equiv {k^3\over2\pi^2} |\delta x|^2 \, ,
\end{equation}
and thus for modes far outside the horizon ($k\eta\to0$) we have
\begin{eqnarray}
\label{pBBphi}
{\cal P}_{\delta\varphi} & = & {32\over\pi^2} l_{\rm Pl}^2\widetilde{H}^2
 (-k\eta)^3[\ln(-k\eta)]^2
 \,,\\
\label{pBBbeta}
{\cal P}_{\delta\beta} & = & {32\over \pi^2} l_{\rm Pl}^2\widetilde{H}^2
 (-k\eta)^3[\ln(-k\eta)]^2
 \,,
\end{eqnarray}
where
$\widetilde{H}\equiv\widetilde{a}'/\widetilde{a}^2=1/(2\widetilde{a}\eta)$
is the Hubble rate in the Einstein frame.
The amplitude of the perturbations grows towards small
scales, but only becomes large for modes outside the horizon
($|k\eta|<1$) when $\widetilde{H}^2\sim l_{\rm Pl}^{-2}$, i.e., the
Planck scale in the Einstein frame.
The spectral tilt of the perturbation spectra is given by
\begin{equation}
\label{specindex}
n-1 \equiv \Delta n_x =  {d\ln{\cal P}_{\delta x} \over d\ln k}
\end{equation}
which from Eqs.~(\ref{pBBphi}) and~(\ref{pBBbeta}) gives $\Delta
n_\varphi=\Delta n_\beta=3$ (where we neglect the logarithmic
dependence).  This of course is the same steep blue spectra we obtained
earlier for the metric perturbations, which of course is
far from the observed near H-Z scale invariant spectrum. We have recently
examined the case of the evolution of the field perturbations in the
non-singular cosmologies of section five and as with the metric-perturbation
case, amongst a number of new features that emerge
there is a slight shift produced in the spectral index
\cite{CarHuaCop01}.

While the dilaton and moduli fields evolve as massless minimally
coupled scalar fields in the Einstein frame, the axion field's kinetic term
still has a non-minimal coupling to the dilaton field. This is evident in
the equation of motion, Eq.~(\ref{pBBdsigmaeom}), for the axion field
perturbations $\delta\sigma$. The non-minimal coupling of the axion to the dilaton
leads to a significantly different evolution to
that of the dilaton and moduli perturbations.

After some algebra, we find that the late time evolution
in this case is logarithmic with respect
to $-k\eta$, (for $\mu\neq0$)
\begin{equation}
\label{pBBsigma}
{\cal P}_{\delta\sigma} = 64\pi l_{\rm Pl}^2 C^2(\mu) \left( {e^{-\varphi}\widetilde{H}
 \over 2\pi} \right)^2 (-k\eta)^{3-2\mu} \, ,
\end{equation}
where $\mu\equiv|\sqrt{3}\cos \xi_*|$ and the numerical coefficient
\begin{equation}
\label{Cofr}
C(\mu) \equiv {2^\mu\Gamma(\mu) \over 2^{3/2}\Gamma(3/2)} \, ,
\end{equation}
approaches unity for $\mu\to3/2$.

The key result is that the spectral index can differ significantly
from the steep blue spectra obtained for the dilaton and moduli fields
that are minimally coupled in the Einstein frame. The spectral index for the
axion perturbations is given by \cite{CopEasWan97,CopLidWan97}
\begin{equation}
\label{Dnsigma}
\Delta n_\sigma = 3 - 2\sqrt{3}|\cos\xi_*|
\end{equation}
and depends crucially upon the evolution of the dilaton, parameterised
by the value of the integration constant $\xi_*$. The spectrum becomes
scale-invariant as
$\sqrt{3}|\cos\xi_*|\to3/2$, which if we return to the
higher-dimensional underlying theory corresponds to a fixed
dilaton field in ten-dimensions.
The lowest possible value of the spectral
tilt $\Delta n_\sigma$ is $3-2\sqrt{3}\simeq-0.46$ which is obtained
when stable compactification has occurred and the moduli field $\beta$
is fixed. The more rapidly the internal dimensions evolve, the steeper
the resulting axion spectrum until for $\cos\xi_*=0$ we have $\Delta
n_\sigma=3$ just like the dilaton and moduli spectra.

When the background axion field is constant these perturbations,
unlike the dilaton or moduli perturbations, do not affect the scalar
metric perturbations. Axion fluctuations correspond to isocurvature
perturbations to first-order. However, if the axion field does
affect the energy density of the universe at later times (for
instance, by acquiring a mass) then the spectrum of density
perturbations need not have a steeply tilted blue spectrum such as
that exhibited by the dilaton or moduli perturbations. Rather, it
could have a nearly scale-invariant spectrum as required for
large-scale structure formation. Such an exciting possibility has received
a great deal of attention recently, notably in
\cite{DurGasSak98,DurGasSak99,MelVerDur99,VerMelDur00,Enqvist01},
and could be a source
for the `curvaton' field recently introduced by Lyth and Wands as a way
of converting isocurvature into adiabatic perturbations \cite{LytWan02}. Time
will tell if the axion has any role to play in cosmological density perturbations.

\section{Smoking Guns?}
Are there any distinctive features that we should be looking out for which would
act as an indicator that the early Universe underwent a period of kinetic driven inflation?
We have already mentioned the possibility of observing the presence of axion fluctuations
in the cosmic microwave background anisotropies. Some of the other smoking guns include:
\begin{itemize}
\item
The spectrum of primordial gravitational waves steeply growing on
short scales, with a spectral index $n_T=3$, although of no
interest on large scales, such a spectrum could be observed by the
next generation of gravitational wave detectors such as the Laser
Interferometric Gravitational Wave Observatory (LIGO) if they are
on the right scale \cite{AllBru97,Maggiore99,CarCopGas}. The
current frequency of these waves depends on the cosmological
model, and in general we would require either an intermediate
epoch of stringy inflation, or a low re-heating temperature at the
start of the post-big bang era~\cite{CopLidLid98} to place the
peak of the gravitational wave spectrum at the right scale.
Nonetheless, the possible production of high amplitude
gravitational waves on detector scales in the pre--big bang
scenario is in marked contrast to conventional inflation models in
which the Hubble parameter decreases during inflation.
\item
Because the scalar and tensor metric perturbations obey the same
evolution equation, their amplitude is directly related. The amplitude
of gravitational waves with a given wavelength is commonly
described in terms of their energy density at the present epoch. For
the simplest pre--big bang models this is given in terms
of the amplitude of the scalar perturbations as
\begin{equation}
\Omega_{\rm gw} = {2\over z_{\rm eq}} {\cal P}_\zeta
\end{equation}
where $z_{\rm eq}=24000\Omega_oh^2$ is the red-shift of
matter-radiation equality. The advanced LIGO configuration will be
sensitive to $\Omega_{\rm gw}\approx10^{-9}$ over a range of scales
around 100Hz. However, the maximum amplitude of gravitational waves on
these scales is constrained by limits on the amplitude of primordial
scalar metric perturbations on the same scale~\cite{CopLidLid98}. In
particular, if
the fractional over-density when a scalar mode re-enters the horizon during
the radiation dominated era is greater than about $1/3$, then that
horizon volume is liable to collapse to form a black hole with
a lifetime of the order the Hubble time and this
would be evaporating today!
If we find PBH's and gravitational waves together then this would
indeed be an exciting result for string cosmology!
\item
Evidence of a primordial magnetic field could have an interpretation in terms
of string cosmology. In string theory
the dilaton is automatically coupled to the electromagnetic field
strength, for example in the heterotic string effective action the photon field
Lagrangian is of the form
\begin{equation}
\label{Lem}
{\cal L} = e^{-\varphi} F_{\mu\nu} F^{\mu\nu} \,,
\end{equation}
where the field strength is derived from the vector potential,
$F_{\mu\nu}=\nabla_{[\mu}A_{\nu]}$.

Now in an isotropic FRW
cosmology the magnetic field must vanish to zeroth-order, and thus
the vector field perturbations are gauge-invariant
and we can neglect the metric back-reaction to first-order.
In the radiation gauge ($A^0=0$, $A^i_{|i}=0$) then the
field perturbations can be treated as vector perturbations on the
spatial hypersurfaces. The field perturbation $A_i$ turns out to have a
clear unique dependence on the dilaton field. In fact the time
dependence of the dilaton (rather than the scale factor) leads to
particle production during the pre--big bang from an initial
vacuum state~\cite{LemLem95,GasGioVen95a,GasGioVen95b}. Using
the pre--big bang solutions given in
Eqs.~(\ref{dila})--(\ref{dilbeta}), we find that the
associated Power spectrum of the gauge fields have a minimum
tilt for the spectral index
for $\xi_*=0$ when $\mu=(1+\sqrt{3})/2$ with a spectral
tilt $\Delta n_{\rm em}=4-\sqrt{3}\approx2.3$. This is still
strongly tilted towards smaller scales, which currently is too
steep to be observably acceptable.
\end{itemize}

\section{Dilaton-Moduli cosmology including a moving five brane.}
We turn our attention in this final section to M-theory, and in
particular to
cosmological solutions of four-dimensional effective
heterotic M-theory with a moving five-brane, evolving dilaton
and $T$ modulus \cite{CopGraLuk01}. It turns out
that the five-brane generates
a transition between two asymptotic rolling-radii solutions, in a manner
analogous to the case of the NS-NS axion discussed in section three.
Moreover, the five-brane motion generally drives the solutions towards
strong coupling asymptotically. The analogous solutions to
those presented in the pre-big-bang involves
a negative-time branch solution which ends in a brane collision
accompanied by a small-instanton transition. Such an exact solution should
be of interest bearing
in mind the recent excitement that has been generated
over the Ekpyrotic Universe scenario, which involves solving
for the collision of two branes \cite{ekpyrotic,ekpyrotic1}.

The four-dimensional low-energy effective theory we will be using
is related to the underlying heterotic M-theory. Of particular
importance for the interpretation of the results is the relation
to heterotic M-theory in five dimensions, obtained from the
11-dimensional theory by compactification on a Calabi-Yau
three-fold. This five-dimensional theory provides an explicit
realisation of a brane-world. The
compactification of 11 dimensional Horava-Witten theory, that is
11-dimensional supergravity on the orbifold $S^1/Z_2\times
M_{10}$, to five dimensions on a Calabi-Yau three fold, leads to
the
appearance of extra three-branes in the five-dimensional effective
theory. Unlike the ``boundary'' three-branes which are stuck to
the orbifold fix points, however, these three-branes are free to
move in the orbifold direction, and this leads to a fascinating new
cosmology.

Our starting point is the four dimensional action
\begin{equation}
 S = -\frac{1}{2\kappa_P^2}\int d^4 x\sqrt{-g}
 \left[\frac{1}{2}R+\frac{1}{4}(\nabla \varphi)^2
 +\frac{3}{4} (\nabla \beta)^2 +\frac{q_5}{2}e^{(\beta -\varphi)}(\nabla z)^2 \right] \; ,
\end{equation}
where $\varphi$ is the effective dilaton in four dimensions, $\beta$ is the size of the orbifold,
$z$ is the modulus representing the position of the five brane and satisfies $0<z<1$, and $q_5$
is the five brane charge.
Due to the non-trivial kinetic term for
$z$, solutions with exactly constant $\varphi$ or $\beta$ do not exist as soon
as the five-brane moves. Therefore, the evolution of all three fields
is linked and (except for setting $z=$ const) cannot be truncated
consistently any further. Looking for cosmological solutions
for simplicity, we assume the three-dimensional
spatial space to be flat. Our Ansatz then reads
\begin{eqnarray}
ds^2 &=& - e^{2 \nu} d \tau^2 + e^{2 \alpha} d {\bf x}^2  \\
\varphi &=& \varphi(\tau ) \\ \alpha &=& \alpha(\tau ) \\ \beta
&=& \beta(\tau )
\\ z &=& z(\tau)
\end{eqnarray}

The cosmological solutions are given by \cite{CopGraLuk01}
\begin{eqnarray}
 \alpha &=& \frac{1}{3}\ln\left|\frac{t-t_0}{T}\right|+\alpha_0 \\
 \beta &=& p_{\beta ,i}\ln\left|\frac{t-t_0}{T}\right|+
        (p_{\beta ,f}-p_{\beta ,i})\ln\left(\left|\frac{t-t_0}{T}\right|^{-\delta}+1
        \right)^{-\frac{1}{\delta}}+\beta_0 \\
 \varphi &=& p_{\varphi ,i}\ln\left|\frac{t-t_0}{T}\right|+
        (p_{\varphi ,f}-p_{\varphi ,i})\ln\left(\left|\frac{t-t_0}{T}\right|^{-\delta}+1
        \right)^{-\frac{1}{\delta}}+\varphi_0 \\
z &=& d \left( 1 + \left| \frac{T}{t- t_0} \right|^{- \delta} \right)^{-1}
+ z_0\; .
\end{eqnarray}
where $t$ is the proper time, the time-scales $t_0$ and $T$ are arbitrary constants
as are the constants $d$ and $z_0$ which parameterise the motion of the five-brane.
For $-\infty < t< t_0$ we are in the positive branch of the solutions and for
$t_0 < t < \infty$ we are in the negative branch.

We see that both expansion powers for the scale factor $\alpha$
are given by $1/3$, a fact which is expected in the Einstein frame.
The initial and final expansion powers for $\beta$ and $\varphi$ are less
trivial and are subject to the constraint
\begin{equation}
 3p_{\beta ,n}^2+p_{\varphi ,n}^2=\frac{4}{3} \label{cons1}
\end{equation}
for $n=i,f$. These are mapped into
one another by
\begin{equation}
 \left(\begin{array}{c}p_{\beta ,f}\\p_{\varphi ,f}\end{array}\right) = P
 \left(\begin{array}{c}p_{\beta ,i}\\p_{\varphi ,i}\end{array}\right)\; ,\qquad
 P = \frac{1}{2}\left(\begin{array}{rr}1&1\\3&-1\end{array}\right)\; . \label{map1}
\end{equation}
This map is its own inverse, that is $P^2=1$, which is
a simple consequence of time reversal symmetry. The power $\delta$
is explicitly given by
\begin{equation}
 \delta = p_{\beta ,i}-p_{\varphi ,i}\; .
\end{equation}
For $\delta <0$ we are in the negative branch and for $\delta >0$ we are in the positive
time branch.
Finally, we have
\begin{equation}
 \varphi_0-\beta_0 = \ln\left(\frac{2q_5d^2}{3}\right)\; .
\end{equation}

The solutions have the following interpretation: at
early times, the system starts in the rolling radii solution
characterised by the initial expansion powers $p_i$ while the five-brane is
practically at rest. When the time approaches $|t-t_0|\sim |T|$ the
five-brane starts to move significantly which leads to an intermediate
period with a more complicated evolution of the system. Then, after a
finite comoving time, in the late asymptotic region, the five-brane
comes to a rest and the scale factors evolve according to another
rolling radii solution with final expansion powers $p_f$. Hence the
five-brane generates a transition from one rolling radii solution into
another one. While there are perfectly viable rolling radii
solutions which become weakly coupled in at least one of the
asymptotic regions, the presence of a moving five-brane always leads
to strong coupling asymptotically, a phenomenon similar to what we
observed in the dilaton-moduli-axion dynamics (see Figure \ref{figflatfrw2}).

\begin{figure}[htb]
    \centering
    \includegraphics[height=12cm]{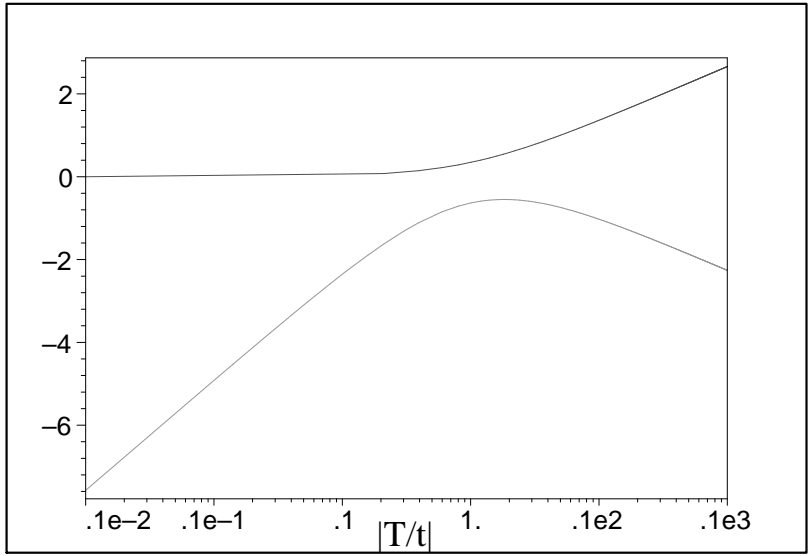}
    \vspace*{-3cm}
    \caption{Time-behaviour of $\beta$ (upper curve)
        and $\varphi$ (lower curve).}
    \label{betaphi}
\end{figure}

\begin{figure}[htb]
    \centering
    \includegraphics[height=12cm]{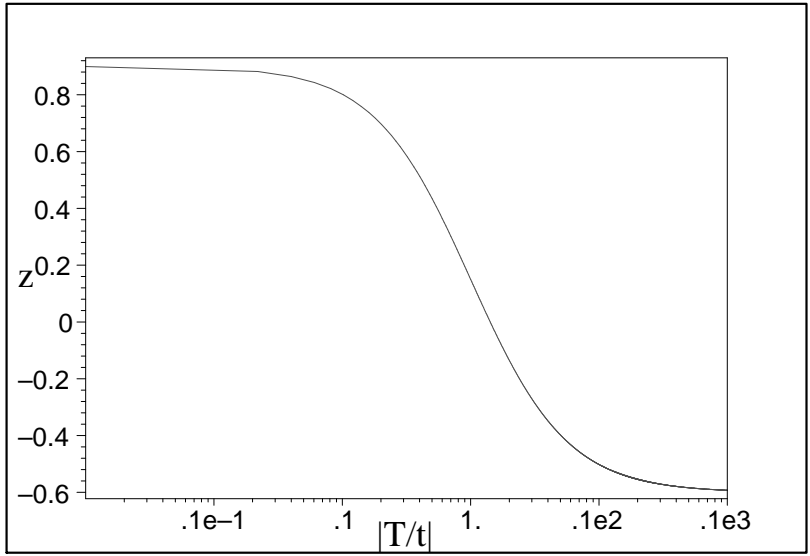}
    \vspace*{-3cm}
    \caption{Time-behaviour of the five-brane position
        modulus $z$ for the example specified in the text.
        The boundaries are located at $z=0,1$ and the five-brane
        collides with the $z=0$ boundary at $|t/T|\simeq 1$.}
    \label{brane}
\end{figure}

These general results can be illustrated by an explicit example.
Focusing on the negative-time branch and considering the solutions
with an approximately static orbifold at early time, Figure
\ref{betaphi} shows the evolution of $\beta$ and $\varphi$,
whereas Figure \ref{brane} shows the evolution of the dynamical
brane.

At early times, $|t-t_0|\gg |T|$,  the evolution is basically of
power-law type with powers $p_i$, because
at early time the five-brane is effectively frozen at $z\simeq d+z_0$
and does not contribute a substantial amount of kinetic energy. This
changes dramatically once we approach the time $|t-t_0|\sim |T|$.  In a
transition period around this time, the brane moves from its original
position by a total distance $d$ and ends up at $z\simeq z_0$. At the
same time, this changes the behaviour of the moduli $\beta$ and $\varphi$
until, at late time $|t|\ll |T|$, they correspond to another rolling
radii solution with powers controlled by $p_f$. Concretely, the
orbifold size described by $\beta$ turns from being approximately
constant at early time to expanding at late time, while the Calabi-Yau
size controlled by $\varphi$ undergoes a transition from expansion to
contraction. We also find that as with the axion case discussed earlier,
the solution runs into strong coupling in both
asymptotic regions $t-t_0\rightarrow -\infty$ and $t-t_0\rightarrow 0$
which illustrates our general result.

In Fig.~\ref{brane} we have shown a particular case
which leads to brane collision. The five-brane is initially located at
$d+z_0\simeq 0.9$ and moves a total distance of $d=1.5$
colliding with the boundary at $z=0$ at the time $|t-t_0|/|T|\simeq 1$.

This represents an explicit example of a negative-time branch solution
which ends in a small-instanton brane-collision. Solving for these systems
has only just the begun, but already interesting features have emerged including
a new mechanism for baryogenesis arising from the
collision of two branes \cite{baryogenesis}.

\section{Summary}
In this article we have addressed a number of issues relating to
string cosmology. We have seen how rolling radii solutions
associated with the low energy string action lead to new
inflationary solutions, and how the inclusion of the axion field
perturbations can generate scale invariant density fluctuations,
although they are primarily isocurvature in nature. The thorny
issues of initial conditions and Graceful Exit facing the pre Big
Bang scenario have been discussed and possible resolutions
proposed. Observational features of string
cosmology today have been discussed including gravitational wave
detection and anisotropies in the cosmic microwave background.
Finally, we have related these solutions to the exciting new
solutions arising in M-theory cosmology, and showed how a moving
five brane could act in a manner similar to the axion field in the
pre Big Bang case. This is an exciting time for string and
M-theory cosmology, the subject is developing at a very fast rate,
and no doubt there will be new breakthroughs emerging over the
next few years. Hopefully out of these we will be in a position to
address a number of the issues I have raised in this article, as
well as other key ones such as stabilising the dilaton and
explaining the current observation of an accelerating Universe.

\Acknowledgements

I am very grateful to the organisors for inviting me to this
wonderful meeting. Following my meeting with him, I would also like to thank Father
Christmas for bringing the presents my daughters had asked for.



\end{document}